# The Landscape of AI in Science Education: What is Changing and How to Respond

Xiaoming Zhai[1], Kent Crippen[2]

[1]*AI4STEM Education Center, University of Georgia*

[2]*School of Teaching and Learning, University of Florida*

This introductory chapter explores the transformative role of artificial intelligence (AI) in reshaping the landscape of science education. Positioned at the intersection of tradition and innovation, AI is altering educational goals, procedures, learning materials, assessment practices, and desired outcomes. We highlight how AI-supported tools—such as intelligent tutoring systems, adaptive learning platforms, automated feedback, and generative content creation—enhance personalization, efficiency, and equity while fostering competencies essential for an AI-driven society, including critical thinking, creativity, and interdisciplinary collaboration. At the same time, this chapter examines the ethical, social, and pedagogical challenges that arise, particularly issues of fairness, transparency, accountability, privacy, and human oversight. To address these tensions, we argue that a Responsible and Ethical Principles (REP) framework is needed to offer guidance for aligning AI integration with values of fairness, scientific integrity, and democratic participation. Through this lens, we synthesize the changes brought to each of the five transformative aspects and the approaches introduced to meet the changes according to the REP framework. We argue that AI should be viewed not as a replacement for human teachers and learners but as a partner that supports inquiry, enriches assessment, and expands access to authentic scientific practices. Aside from what is changing, we conclude by exploring the roles that remain uniquely human—engaging as moral and relational anchors in classrooms, bringing interpretive and ethical judgement, fostering creativity, imagination, and curiosity, and co-constructing meaning through dialogue and

---

community—and assert that these qualities must remain central if AI is to advance equity, integrity, and human flourishing in science education.

## 1. Science Education is at a Crossroads

Science education currently stands at a pivotal crossroads, increasingly shaped by the rapid convergence of longstanding pedagogical traditions alongside the rise of a disruptive innovation—artificial intelligence (AI). AI is no longer confined to isolated industries; its transformative impact now permeates virtually every sector of society such as healthcare, finance, transportation, manufacturing, communication, and including science education (Guo, Zheng, et al., 2024; Lee et al., 2025). These widespread changes are altering the nature of scientists' work yet are also redefining how scientific knowledge is generated, spread, and applied (Herdiska & Zhai, 2024). Scientists are leveraging AI tools to expedite complex research processes such as data mining, pattern recognition, and simulation in fields like genomics, neuroscience, and environmental science (Eraslan et al., 2019; Gil et al., 2014). This technological evolution is having a cascading effect on the workforce, necessitating new competencies that extend far beyond traditional disciplinary knowledge (Shi, 2025). Employers increasingly value critical thinking, data literacy, adaptability, and interdisciplinary collaboration—skills that are often underdeveloped in conventional educational frameworks (Cheng et al., 2025; OECD, 2021). As the societal fabric becomes more intertwined with AI capabilities, science education must evolve in parallel to prepare learners not only to participate in this new landscape but also to lead within it.

These far-reaching societal and occupational shifts demand a fundamental rethinking of science education to ensure that students emerge as informed, capable contributors in both civic and professional realms. AI's integration into daily life intensifies this need, exposing the inadequacy of educational systems still rooted in standardized science curricula and outdated assessment practices. For over a decade, researchers have drawn attention to persistent limitations in traditional science education, including declining student engagement, limited differentiation, and a troubling disconnection between classroom instruction and real-world application (OECD, 2017; Zhai & Pellegrino, 2023). Moreover, structural inequities continue to manifest through significant achievement gaps along socioeconomic lines, as many under-resourced schools lack access to current materials, qualified educators, and interactive learning environments (Hu & Morgan, 2024). When viewed against the backdrop of an AI-driven society, these deficits signal an urgent need for systemic reform—one that prioritizes equity, responsiveness, and the cultivation of future-ready skills through science education (Varsik & Vosberg, 2024). Building on this need for transformation, AI is increasingly recognized as a driving force capable of bridging the gap between outdated science education models and contemporary societal demands.

Positioned as a catalyst for innovation, AI introduces a suite of tools and frameworks that directly confront long-standing challenges in science education. These technologies include AI-supported scoring of open-ended written explanations and arguments or drawn models that delivers rubric-aligned scores and formative feedback at scale (Zhai., 2024), allowing students to practice modeling, explaining, and arguing from evidence while teachers track growth with reliability approaching human ratings. Adaptive learning platforms that tailor instruction to individual students' cognitive and emotional profiles, promote differentiated learning experiences that foster deeper understanding (Gao & Zhang, 2025). Intelligent Tutoring Systems (ITS), such as AutoTutor and EER-Tutor, replicate aspects of one-on-one instruction by offering real-time, customized feedback based on student inputs, thereby enhancing engagement and learning efficiency (Latif et al., 2025). Generative AI further enables automatic creation of instructional materials, such as lesson plans and assessments (Lee & Zhai, 2024), reducing teachers' routine workload and freeing them to concentrate on higher-order instructional tasks that more directly support science learning. In addition, AI-supported learning analytics can provide educators with fine-grained, real-time insights into student behavior, comprehension levels, and progression, enabling data-informed instructional decisions for science learning (Gašević et al., 2022; Sajja et al., 2025). These analytics help identify learning bottlenecks, predict at-risk students, and support timely interventions. Empirical studies show that AI-supported platforms like the Web-based Inquiry Science Environment (WISE) (https://wise.berkeley.edu/ ) significantly improve academic performance, especially among students who benefit from personalized pacing and targeted scaffolding (Li et al., 2024). Through such capabilities, AI is not merely supporting existing science pedagogical strategies but fundamentally re-engineering the teaching and learning process to align with the demands of a future-oriented, AI-driven society. Science education is at a crossroads due to the significant ongoing changes brough by AI, and understanding these changes better equips us to respond thoughtfully and productively.

## 2. An Educational Transformation Model in the Era of AI

Extending the impact of AI technologies beyond tools and platforms, these innovations are fundamentally altering the aims and structure of science education. Educational goals, the broad, guiding purposes that articulate what education seeks to achieve, are shifting from rote memorization toward fostering essential twenty-first-century scientific competencies that ensure graduates to be succeed in AI environments (Zhai, 2023). With the integration of AI-supported learning environments to facilitate instructional activities, students can feasibly engage in science and engineering practices that cultivate disciplinary core ideas and crosscutting concepts. AI can enhance the development of critical thinking and inquiry-based reasoning by enabling adaptive simulations, intelligent tutoring, and real-time feedback that personalize exploration (Mafarja et al., 2025). Students build AI literacy by learning to navigate, interpret, and critically assess AI-supported tools, algorithms, and data systems (Cheng et al., 2025). These changes align with broader societal needs for a scientifically literate workforce capable of tackling interdisciplinary challenges in biotechnology, climate science, energy systems, and data-intensive research (National Research Council, 2012). By aligning goals with the realities of an AI-driven society, science education is moving toward preparing students not only to understand scientific content but also to apply scientific knowledge to solve pressing global problems.

The transformation of *educational goals* in science education reflects a paradigm shift in what it means to be scientifically literate. Scientists are increasingly relying on AI for significant discoveries (Bianchini et al., 2022; Kitano, 2016), and these innovations need to be introduced to students if we expect them to succeed in contemporary science and achieve scientific literacy in the era of AI (Vartiainen et al., 2021). Instead of emphasizing factual recall and algorithmic problem-solving, the focus has shifted toward cultivating higher-order thinking skills, the ability to interpret and analyze scientific data, and the capacity to engage critically with emerging technologies such as AI, gene editing, and the ever expanding applications of computational modeling (Sinha et al., 2023). Students are also being prepared to grapple with the ethical and societal implications of scientific innovation, ensuring that they can act as responsible citizens and professionals in a rapidly changing world (Akgün & Krajcik, 2024b). This transformational goal situates science education at the core of equipping learners with the intellectual agility required for lifelong learning and cross-disciplinary collaboration.

Equally important is the transformation of *educational procedures* in science classrooms, which are being redesigned to promote inquiry, experimentation, and collaborative problem-solving. Educational procedures refer to the structured methods, routines, and pedagogical practices educators use to organize and deliver instruction. With the integration of AI technologies, pedagogical models such as flipped classrooms, problem-based learning, and inquiry-driven laboratory work are continuing to gain in relevance, and prominence. AI-supported platforms provide real-time feedback, simulate experimental conditions, and personalize learning sequences, enabling students to design and test hypotheses, analyze experimental data, and refine their understanding of scientific concepts (Krenn et al., 2022; Lee et al., 2024b). More importantly, the instructional and learning process can be more efficient with the support of AI. For example, intelligent tutoring systems can guide students through complex topics such as molecular dynamics or climate modeling, offering targeted support where misconceptions arise and scaffolding deeper conceptual growth (Gobert et al., 2024). This transformation inevitably requires teachers to develop new knowledge and skills to accommodate the changes (Shi & Choi, 2024), thus improving the overall learning experience for students.

*Learning materials* in science education, the resources designed or adapted to support the learning process, are also undergoing substantial transformation, increasingly evolving into interactive, adaptive, and multimodal forms. Traditional static textbooks are being replaced by virtual laboratories, AI-supported simulations, and augmented reality environments that allow students to engage directly with scientific phenomena (Lee et al., 2024a). For instance, platforms such as Google Teachable Machine provide opportunities for learners to manipulate variables and train algorithms for scientific inquiry (Herdliska & Zhai, 2024), while generative AI such as GPT offers opportunities to explain graphical representations for real-world phenomena (Bewersdorff et al., 2025). In addition, to mimic scientists' authentic research, these AI-supported materials promote inclusivity and accessibility by adjusting to individual learning preferences, linguistic backgrounds, and cognitive needs, making advanced scientific exploration available to diverse populations of learners worldwide. Moreover, AI is playing an increased role in assisting educators and researchers in creating up-to-date learning materials (Su & Zhong, 2022), freeing curriculum designers from heavy time-consuming work so that they can focus on more intellectual design and development.

*Assessment practices* in science education, both formative and summative, are similarly being reshaped by AI, moving away from standardized, summative testing toward formative, performance-based, and authentic evaluation (Linn & Gerard, 2024). AI has enabled automatic scoring of students' written explanations and arguments and drawn models, significantly advancing automaticity, and extending the complexity of constructs that assessment can target (Nehm, 2024; Zhai et al., 2020). AI-supported learning analytics can track how students design experiments, interpret data, and communicate scientific arguments, thereby assessing both process and product in scientific inquiry (Pellegrino, 2024). Real-time dashboards for teachers and learners capture patterns of misunderstanding in topics like genetics or thermodynamics, allowing for immediate intervention. In addition, longitudinal assessment models supported by AI generate comprehensive learner profiles, making it possible to trace the development of scientific reasoning skills across years of study (Liu et al., 2024), thereby embedding assessment as a continuous component of the learning cycle rather than as a final endpoint.

Ultimately, these transformations in goals, procedures, materials, and assessment culminate in a redefinition of *educational outcomes* in science education, the knowledge, skills, attitudes, and capacities that learners develop. Outcomes are no longer confined to the mastery of scientific facts or completion of laboratory exercises; they now emphasize creativity in experimental design, collaboration in interdisciplinary contexts, resilience in problem-solving, and ethical reflection on the applications of science and technology, including AI. Students who have mastered AI-supported science education are envisioned as adaptive, lifelong learners who are capable of addressing complex scientific and societal challenges with the responsible and ethical use of AI. This transformation ensures that science education not only conveys disciplinary knowledge but also equips learners with the competencies required to lead innovation in an AI-driven future.

**3. Obstacles Along the Journey**

Yet, despite AI's transformative promise, its integration into science education brings a complex and expansive array of ethical, social, and pedagogical concerns that demand not only closer examination but ongoing critical reflection. At the forefront of these concerns is the issue of fairness (Xing & Li, 2024). While AI holds the potential to enhance learning opportunities equitably, its current deployment may reflect and reinforce systemic inequities already embedded in educational systems. For example, algorithmic bias, stemming from historically skewed, incomplete, or non-representative datasets, can entrench preexisting disparities rather than address them. Educational AI tools, especially those trained predominantly on data from affluent or high-performing schools, may unintentionally marginalize students from under-resourced environments. This marginalization can manifest in various ways—from unfair or inaccurate feedback to faulty predictions that influence academic placement or opportunities, leading to a form of digital gatekeeping that limits student potential and exacerbates existing achievement gaps (Kubsch et al., 2022). In addition, over-cautiousness and pseudo AI bias have emerged as significant factors that prevent the acceptance of AI, leading to expanded disparities among users in different groups (Zhai & Krajcik, 2024). These concerns become especially acute in diverse classrooms where educational equity is already a persistent challenge.

Transparency and explainability present equally significant issues. A troubling number of AI systems function as opaque boxes, generating decisions and recommendations that even developers may struggle to fully interpret (Wu et al., 2024). In educational contexts where AI influences assessment, learning pathways, or instructional resources, students and educators must be able to discern the reasoning behind the system's outputs. When transparency is lacking, the result is a troubling reliance on machine-generated judgments that go unscrutinized, undermining the educational process and placing undue trust in unaccountable algorithms. Furthermore, this opacity leads directly to the issue of accountability (Akgün & Krajcik, 2024a). When AI systems make mistakes—whether through flawed programming, biased data, or contextual misinterpretation—who is held responsible? Unfortunately, educational settings often lack clear protocols for assigning accountability, leaving students and teachers without recourse when AI tools cause harm or misjudgement.

Privacy and data security add another dimension to the ethical landscape. AI applications in education rely heavily on gathering vast amounts of data, including not only academic performance but also behavioral patterns, engagement levels, and sometimes even biometric or location-based information (Latif & Zhai, 2025). This extensive data collection raises urgent questions about consent, control, and usage. For example, who owns this data? How long is it stored? Who has access, and under what circumstances? Increasing concerns have emerged about the exploitation of student data for commercial purposes or surveillance, undermining institutional trust and student autonomy. More than a technical challenge, data privacy reflects a broader institutional value system: whether education remains a protected space for intellectual growth or becomes a site of commodified surveillance.

Concerns about human oversight and autonomy further complicate AI's role in the classroom. While AI can offer valuable support by automating routine tasks or generating real-time analytics, it cannot and should not replace the human dimensions of teaching (Shi & Choi, 2024). There is a growing risk that over-reliance on AI will diminish the educator's role, reducing teachers to monitors of algorithm-driven instruction and students to passive recipients of machine-curated content. This shift could erode opportunities for meaningful dialogue, mentorship, and moral development—hallmarks of quality education. Students must be empowered to question AI recommendations and remain active participants in their learning journey, rather than being conditioned to accept technological authority as absolute. The erosion of human judgment and agency in classrooms poses a serious threat to the core educational mission of fostering independent, critical thinkers.

Another pressing concern involves the intended purposes behind AI use in educational environments. While AI is often introduced under the guise of innovation, many implementations serve efficiency-driven objectives, such as standardizing content delivery or streamlining testing procedures. Such applications, without careful attention and thoughtful human-involvement, risk reducing science education to mechanized instruction, stripped of the exploratory, creative, and inquiry-based elements that make scientific learning meaningful. This could be particularly detrimental to historically minoritized populations in STEM fields, for whom sustained engagement and personal relevance are essential for inclusion and long-term success (Nyaaba et al., 2024). The tension between genuine educational enrichment and superficial AI integration calls for deep reflection on whose needs are being prioritized and what pedagogical values are being upheld.

Further complicating the integration of AI into science education are concerns surrounding robustness and scientific integrity. Despite bold claims of precision and adaptability, AI systems remain susceptible to technical limitations. Inaccurate models, poorly curated datasets, or oversimplified algorithms can distort complex scientific concepts or propagate misinformation or feedback (e.g., over inferences or over praise, see Guo, Latif, et al. (2024)). The iterative and often ambiguous nature of scientific inquiry—which depends on uncertainty, debate, and evolving evidence—can be flattened by AI systems that privilege deterministic outputs. At worst, students may receive misleading or overly confident explanations that contradict best practices in scientific reasoning. Such distortions undermine science education's foundational commitment to evidence-based thinking and intellectual rigor.

Lastly, there is an overarching concern that AI could erode essential human values in education. As systems increasingly quantify and rank learners through data analytics, there is a risk of reducing students and teachers to mere data points. This process may inadvertently marginalize diverse cultural perspectives and suppress ethical discourse in favor of standardized metrics. Moreover, embedded biases in AI systems may reflect dominant ideologies, leading to exclusions that are subtle yet powerful in shaping who feels seen, valued, and heard within educational spaces. These dynamics raise fundamental questions about moral agency, ethical responsibility, and the kind of society education aims to cultivate.

Without sustained and critical attention to these multifaceted concerns—including fairness, transparency, accountability, privacy, human oversight, engagement, robustness, and respect for human values—AI in science education risks becoming a force of simplification and exclusion rather than empowerment and inclusion. Instead of enhancing learning, it may deepen existing divides, reduce science to algorithmic exercises, and erode the human relationships and reflective capacities that lie at the heart of meaningful education.

**4. Our Approach: Responsible and Ethical Principles for the Practice of AI-Supported Science Education**

The concerns raised in the previous section—*fairness, transparency, accountability, privacy, human oversight, and the preservation of human valu*es—make clear that the integration of AI into science education must be guided by more than technical feasibility or efficiency. Left unchecked, AI could deepen inequities, obscure accountability, or diminish the agency of teachers and students. At the same time, its potential for supporting inquiry, enabling personalization, and widening access to scientific practices is undeniable. Navigating these tensions requires an intentional framework that balances innovation with integrity. It is not enough to ask what AI *can* do in classrooms; we must also ask *how* it should be used and *why* it should be used at all.

To meet this challenge, we turn to the Responsible and Ethical Principles (REP) Framework, which is developed more fully later in the book (See Chapter 3, Crippen et al., 2025). The REP framework makes an important distinction between *responsible principles*—those that focus on the safeguards, practices, and methods for implementing AI (i.e., how)—and *ethical principles*—those that clarify the purposes, values, and broader human aims that justify its use (i.e, why). By holding these two dimensions together, the framework provides a balanced approach: it insists that AI systems be transparent, fair, and accountable, while also reminding us that their ultimate purpose is to promote meaningful learning, uphold scientific integrity, and respect the dignity and diversity of all learners.

In this book, the REP framework is used not as a compliance checklist but as a lens for educational transformation. Applied across the five core areas identified in our model—goals, procedures, materials, assessment, and outcomes —REP helps us see how AI can reshape science education in ways that align practice with values. For example, it frames educational goals around cultivating inquiry and critical reasoning rather than passive knowledge acquisition; it encourages classroom procedures that use AI to amplify rather than replace teacher and student agency; it calls for learning materials that are not only adaptive but also inclusive and culturally relevant; it informs assessment practices that prioritize authenticity and ownership of learning; and it guides the envisioning of outcomes that prepare learners not only to use AI tools but also to critique and shape them as informed citizens and professionals. In this way, REP links ethical commitments with practical action across the entire educational system. Our approach, then, positions AI as a partner in science education rather than a replacement for human judgment or relational teaching. By embedding REP into the design and use of AI, we ensure that innovation does not erode the values that make science education meaningful: equity, inquiry, creativity, integrity, and democratic participation. This orientation invites educators, students, policymakers, and designers to see AI not as a neutral tool but as a technology whose influence must be continually interpreted, critiqued, and guided by shared human purposes.

Below, we weave the eight principles of the REP into a synthesized call for action on how they may unfold in science education (for details see chapter 3, Crippen, Zhai, & Lee, 2025):

*Fairness in Opportunity and Design*. Science education must ensure that AI expands, rather than restricts, participation. Fairness means more than avoiding overt bias; it requires proactive inclusion. Imagine an AI simulation platform for physics that disproportionately recommends advanced challenges to students whose prior performance signals mastery, but neglects learners still developing foundational skills. Such practices could amplify inequities. To counteract this, teachers and developers must design adaptive systems that scaffold struggling learners without narrowing their opportunities. Inclusive design can be seen in AI-supported lab simulations that provide multilingual scaffolds or accessibility features for students with disabilities, thereby broadening participation in authentic scientific inquiry.

*Protecting Privacy and Data Security.* As AI systems rely heavily on learner data, protecting privacy is both a technical and ethical obligation. Data should not only comply with Family Educational Rights and Privacy Act (FERPA) and the General Data Protection Regulation (GDPR). but also align with contextual sensitivities in the classroom. For example, an AI dashboard tracking student progress should present insights without exposing sensitive personal details to peers. Teachers can model ethical reasoning by engaging students in discussions about consent and data use, making privacy a topic of scientific and civic learning rather than a hidden technical concern. Robust data protections reassure learners that participation in science education does not come at the cost of their dignity or autonomy.

*Affirming Human Oversight and Autonomy*. AI should never displace the expertise of teachers or the agency of learners. For instance, when AI generates a step-by-step solution to a chemistry problem, the teacher's role is to challenge students to analyze, critique, or extend that solution rather than accept it passively. By situating AI as a partner for exploration, learners develop critical stances toward evidence and reasoning. Similarly, teachers should retain decision-making authority in adapting AI-supported recommendations about instruction, ensuring that the unique social and relational dimensions of classrooms are preserved.

*Fostering Beneficial and Meaningful Use.* AI must be deployed with a focus on enriching authentic learning rather than streamlining efficiency alone. In practice, this may mean leveraging AI to support inquiry-based projects where students use AI-supported visualizations of ecological data to form hypotheses and test them through field investigations. Such applications turn AI into a catalyst for curiosity and deeper reasoning, enabling learners to engage more directly in the practice of science. By contrast, systems that automate grading without meaningful feedback risk diminishing the learning process and reducing opportunities for reflection.

*Upholding Robustness and Scientific Integrity.* Scientific accuracy and reliability are non-negotiable in AI-supported learning. Consider a scenario in which an AI chatbot explains photosynthesis but provides oversimplified or inaccurate causal mechanisms. Left unchecked, such outputs undermine scientific literacy. Developers must rigorously test systems for alignment with disciplinary knowledge, and teachers must cultivate classroom practices where AI outputs are treated as hypotheses to be tested, not facts to be memorized. In doing so, learners develop both epistemic humility and scientific reasoning skills.

*Respecting Human Values and Cultural Identities.* Science education must foreground the dignity and cultural identities of learners. AI tools can be designed to incorporate culturally responsive contexts—for example, presenting climate change data using local environmental issues that resonate with students' lived experiences. Teachers can also use AI-supported prompts to facilitate classroom debates about the ethical implications of emerging technologies, thereby strengthening both civic reasoning and cultural sensitivity. In this way, AI enhances, rather than erodes, the human dimensions of science education.

*Promoting Transparency and Explainability.* Complex AI models must be rendered interpretable for teachers and learners. A biology class might use an AI tool that predicts species survival under different climate conditions. Rather than treating predictions as opaque, students should be given insights into the data inputs, assumptions, and limitations behind the model. Teachers can frame these explanations as opportunities for critical engagement with uncertainty and scientific modeling, thereby strengthening data literacy. Transparency allows learners to not only consume AI outputs but also interrogate how those outputs are produced.

*Embedding Accountability Across Stakeholders.* Accountability ensures that responsibility rests with the humans and institutions that design and implement AI, not with learners or the technology itself. For example, if an AI-supported assessment system consistently underrates the performance of students from certain linguistic backgrounds, schools must have clear mechanisms for appeal, review, and correction. Developers, too, must commit to monitoring long-term impacts and communicating failures openly. This shared accountability reinforces trust and ensures that AI use remains aligned with educational goals.

**5. Overview of the Book: What AI is Changing and How Should We Respond**

In response to these complex dynamics, in this book, the authors undertake a comprehensive analysis of how AI is reshaping the landscape of science education. We critically examine how AI is catalyzing shifts in educational goals toward scientific literacy, such as critical thinking, ethical reasoning, and interdisciplinary problem-solving. We explore the transformation of teaching practices through intelligent systems that support differentiated instruction and real-time scaffolding, and the redesign of curriculum to include AI literacy, data science, and inquiry-driven exploration. We also analyze changes in assessment modalities, highlighting the move from static, summative evaluations to dynamic, formative assessments informed by real-time analytics and predictive feedback. Finally, the authors investigate how these changes collectively redefine learning outcomes, emphasizing the development of agile, collaborative, and reflective learners equipped to thrive in an AI-driven future. Drawing on empirical evidence and case studies, we propose ethically grounded, practically viable strategies for integrating AI in ways that amplify educational equity, foster scientific literacy, and prepare learners for the evolving demands of the twenty-first century. In the subsequent section, we outline the key ideas for each chapter of the book according to the components of educational transformation: educational goals, educational procedures, learning materials, assessment, and educational outcomes.

### *5.1. Educational Goals*

To address the profound changes to science education goals brought by AI, this book includes two dedicated chapters on the topic. *Chapter 3* highlights how AI is reshaping science literacy by expanding it across three visions. In *Vision I*, AI transforms core scientific practices such as hypothesis generation, modeling, data analysis, and solution design, requiring students to develop competencies in AI-supported scientific inquiry rather than focusing solely on traditional content mastery. In *Vision II*, AI operates as a collaborative partner in applying science to solve authentic, real-world problems, urging learners to engage with AI tools critically and creatively. *Vision III* extends science education goals further by empowering students as agents of change, where responsible and ethical engagement with AI becomes crucial for addressing socio-political, cultural, and environmental challenges. Collectively, these changes call for an integrative model of *Science Literacy Plus*, where AI literacy is interwoven with science literacy to prepare learners for active participation in an AI-supported world.

To meet these evolving goals responsibly and ethically, Chapter 3 stresses the need to embed ethical frameworks and critical reflection into AI-supported science education. Responsible practices include ensuring equitable access to AI tools, mitigating algorithmic bias, and fostering transparency, accountability, and human oversight. Educators are encouraged to guide students in recognizing AI's limitations, questioning whose perspectives are represented, and reflecting on the broader social and environmental impacts of technological applications. Ethical engagement also involves protecting privacy, respecting Indigenous knowledge, and promoting sustainability in the development and use of AI systems. Ultimately, we argue that science education should cultivate not only technical proficiency but also values of empathy, justice, and collective responsibility, enabling learners to leverage AI for societal good while avoiding harm.

*Chapter 4* builds on these insights by emphasizing that AI is fundamentally transforming science education goals from memorization and routine experimentation to sustained inquiry and lifelong investigation. By combining the advances brought by AI with Knowledge Integration (KI) pedagogy, classrooms can become design laboratories where students engage in authentic investigations, interpret complex phenomena, and devise solutions to pressing societal and environmental challenges. In this case, AI scaffolds the inquiry cycle by supporting students as they brainstorm researchable questions, visualize and test models, analyze data, and refine explanations. This shift reorients educational goals toward agency, creativity, and problem-solving, ensuring learners are prepared not only to understand science but also to apply it in addressing real-world dilemmas alongside AI tools.

To ensure these transformations occur responsibly, Chapter 4 underscores the importance of aligning AI-supported inquiry with principles of fairness, transparency, accountability, and human oversight. Ethical use of AI in science education requires that guidance systems honor the diversity of students' ideas without reinforcing bias, while safeguarding privacy and ensuring data security. Partnerships among teachers, students, and designers are essential to critically examine AI outputs, refine instructional uses, and preserve student agency in investigations. Professional development and collaborative curriculum design are also needed so that educators can effectively mediate AI integration while upholding inclusivity, human dignity, and democratic participation. By embedding responsible practices into AI-supported inquiry, science education can leverage AI's opportunities while mitigating risks, preparing learners as autonomous investigators and ethical participants in an AI-driven society.

In summary, *Chapters 3 and 4* together illustrate a comprehensive transformation of science education goals driven by AI. While Chapter 3 reframes science literacy through the lens of AI across three visions, Chapter 4 emphasizes the inquiry-focused practices that AI can scaffold in classrooms. Both chapters highlight that realizing AI's potential in education requires a strong ethical foundation—one that balances technical advancement with values of fairness, inclusivity, sustainability, and democratic participation. Together, these chapters chart a path toward cultivating learners who are not only scientifically and technologically literate but also grounded in the ethical and socially responsible use of AI.

### *5.2. Educational Procedures*

*Chapters 6* and *7* collectively illuminate the transformative role of AI in reshaping science education by redefining both teacher practice and student learning. These chapters emphasize that the integration of AI is not a peripheral enhancement but a structural shift toward more data-informed, student-centered, and co-orchestrated approaches to teaching and learning.

*Chapter 6* underscores the transformative impact of AI within science teacher education, where these technologies are reshaping teacher practice by streamlining traditionally time-intensive tasks such as lesson planning, formative assessment design, and classroom orchestration. Through natural language processing and machine learning applications, teachers now have access to automated scoring systems, intelligent dashboards, and personalized feedback mechanisms that highlight student misconceptions in real time. This allows educators to adapt instruction more responsively, aligning it more closely with the multidimensional goals of frameworks like the Next Generation Science Standards (NGSS; NGSS Lead States, 2013). Moreover, the use of AI can support teacher professional growth through simulations and practice-based environments where teachers can rehearse core instructional competencies, such as eliciting student thinking or facilitating scientific argumentation, in reduced-complexity contexts. These developments not only save time but also extend opportunities for teachers to refine their craft with ongoing, targeted, and data-informed insights. In sum, AI integration signifies a shift from labor-intensive processes to more efficient, evidence-based, and student-centered instructional decision-making, positioning teachers to more effectively meet the needs of diverse learners while continuing to advance their professional development.

Addressing these changes responsibly and ethically requires a multi-faceted approach that emphasizes equity, transparency, and human-centered practice. The chapter highlights that while AI tools can amplify instructional precision, they also carry risks related to bias, privacy, and inequitable access, particularly in under-resourced educational contexts. Ethical use demands that teachers and school systems critically interrogate how algorithms are trained, what assumptions underlie their design, and how their outputs are interpreted in relation to student learning. Professional development is thus central, not only for building technical fluency but also for fostering educators' ability to question, contextualize, and adapt AI recommendations within culturally responsive and justice-oriented pedagogies. Equitable implementation also requires systemic commitments—such as providing infrastructure, sustained support, and opportunities for collaboration across institutions—to ensure that AI enhances rather than exacerbates existing disparities. Ultimately, the responsible integration of AI hinges on positioning it as a complement to, rather than a substitute for, teachers' expertise, creativity, and ethical judgment. By doing so, schools can harness AI's efficiency and analytical power to expand inquiry, foster deeper engagement, and cultivate classrooms where technological innovation aligns with core values of inclusivity, fairness, and respect for the human dimensions of learning

*Chapter 7* introduces three significant changes to the procedures of science education--enhancing the personalization, authenticity, and efficiency of learning processes. Through predictive and generative technologies, AI systems support teachers in tailoring instruction to diverse learners, provide real-time feedback, and create adaptive learning environments such as intelligent tutoring systems and exploratory learning platforms. These tools enable students to engage more actively and autonomously in inquiry-based learning, while also offering teachers synthesized insights about student performance and needs. In addition, AI-supported platforms can create multiple versions of assignments, provide visualizations, and generate scaffolded supports that address students' varied levels of prior knowledge. This flexibility allows educators to more effectively accommodate the wide range of interests, competencies, and cultural backgrounds present in modern classrooms. Moreover, AI reshapes classroom dynamics by functioning not only as a tool but also as an epistemic partner, thereby redistributing roles of authority and responsibility among teachers, students, and technological agents. As a result, science education is evolving into a more individualized, data-informed, and co-orchestrated learning ecosystem that can better address the complexity of student diversity, the challenges of authentic scientific inquiry, and the broader demands of modern scientific literacy.

The transformative impact of AI in science education procedures, however, underscores the necessity of responsible and ethical integration. To safeguard meaningful learning, educators must design instructional strategies that integrate AI with curricula rather than rely on AI as a stand-alone solution. Policies and frameworks are needed to define clear roles for teachers, students, and AI, ensuring that technology complements rather than replaces human judgment. Ethical challenges such as bias in training data, lack of transparency in algorithmic processes, and issues of inclusivity demand systematic attention, alongside efforts to foster AI literacy among both teachers and students. Professional development programs are essential to equip educators with the competencies required for critical evaluation, effective orchestration, and the nurturing of students' epistemic autonomy. Moreover, institutional support is required to ensure that AI adoption does not widen educational inequities but instead contributes to inclusive and equitable opportunities for learning. Teachers and students must learn to critically interrogate AI outputs, engage in reflective practices, and understand both the potentials and the limitations of algorithmic systems. Ultimately, responsibility rests with human actors to ensure that AI tools amplify equity, fairness, transparency, and deep engagement in science learning rather than reduce education to superficial or automated processes.

### *5.3. Learning Materials*

*Chapter 8* addresses the changes to science learning materials by highlighting AI's profound role in transforming the design, function, and accessibility of the resources used to support learning. AI expands the scope of learning materials beyond static, text-based representations to dynamic, adaptive, and multimodal environments. Unlike conventional textbooks that present fixed facts and limited interactivity, AI-supported platforms integrate authentic scientific practices such as hypothesis testing, large-scale data analysis, and predictive modeling, mirroring how science is conducted in contemporary fields like genomics and climate research. Adaptive systems powered by AI adjust the pace, modality, and complexity of instruction in real time, providing personalized learning trajectories that respond to students' prior knowledge, interests, and cultural backgrounds. Interactive AI simulations, for instance, allow learners to design and manipulate experiments, analyze outcomes, and iteratively refine their models, fostering inquiry-based and evidence-driven learning. At the same time, generative AI facilitates co-creation by enabling teachers to rapidly develop culturally responsive and contextually relevant materials—lesson plans, formative assessments, or analogies to abstract concepts—that evolve alongside both curricular standards and scientific discoveries. AI also plays a critical role in advancing accessibility by providing translation, text simplification, and multimodal adaptations for neurodiverse and multilingual learners, thereby supporting more equitable participation in science classrooms. Taken together, these changes reimagine science education as more participatory, responsive, and reflective of the computational, data-intensive nature of modern science.

Nevertheless, the transformative potential of AI in science learning materials is inseparable from pressing ethical and pedagogical challenges, which demand responsible and deliberate responses. Without safeguards, algorithmic bias, misinformation, and opacity in AI-supported content risk reinforcing educational inequities, misrepresenting science, or undermining student and teacher agency. Ethical integration therefore requires transparency measures such as documentation of AI outputs, explainability features, and opportunities for students to critique, validate, and reflect on AI-supported insights. Privacy protections are equally crucial, as adaptive learning environments depend on sensitive learner data that must be anonymized, secured, and governed by clear consent protocols. Equally important is ensuring scientific integrity: AI-supported simulations and multimodal resources must be grounded in validated models and pedagogical coherence, requiring collaboration among educators, domain experts, and technologists. Accessibility features must be designed with cultural sensitivity and inclusivity in mind, ensuring that AI tools not only translate dominant scientific discourse but also affirm diverse epistemologies and linguistic identities. Moreover, the role of AI as a co-author in curriculum development must remain transparent and accountable, with human educators retaining oversight to ensure alignment with disciplinary knowledge, instructional goals, and ethical norms. By embedding such responsible practices, AI can serve as a collaborative partner that enriches inquiry, creativity, and accessibility while upholding the foundational values of equity, critical reasoning, and scientific integrity in education.

### *5.4. Assessment*

In this section, *Chapters 9* and *10* attend to AI-supported assessments from different perspectives. *Chapter 9* introduces how AI is reshaping science assessment by redefining constructs, introducing new formats, and transforming classroom practices in ways that both expand opportunities and complicate traditional notions of measurement. Traditional assessments, which largely focused on content recall or procedural fluency, are increasingly insufficient in an era where generative AI can provide immediate and often sophisticated answers to many standardized tasks. Instead, AI is pushing assessment design toward performance-based models that emphasize authentic reasoning, complex problem-solving, collaborative inquiry, and human–AI partnerships. These changes include the emergence of AI-resilient assessments that require students to critique, refine, or extend AI-generated outputs, and the use of multimodal data streams—including text, diagrams, gestures, and speech—to capture complex learning processes. AI also enables new possibilities for formative assessment through real-time feedback, adaptive diagnostics, and integration with classroom simulations that mirror professional scientific practice. Importantly, these innovations make it possible to trace student learning trajectories over time, revealing misconceptions, incremental conceptual changes, and opportunities for targeted intervention. Together, these transformations move science assessment away from static measurement and toward dynamic, iterative, and context-sensitive evaluations that better reflect the nature of contemporary science learning.

The chapter further argues that responsibly and ethically addressing these changes requires principled approaches that safeguard validity, fairness, and equity, while also supporting instructional utility. Developers must ensure that AI-scoring systems are transparent, interpretable, and consistently aligned with assessment constructs, rather than producing black-box judgments. Human-in-the-loop models are essential to preserve teacher oversight, while data sampling strategies must be carefully designed to mitigate bias across diverse student populations and to ensure cultural and linguistic inclusivity. Ethical assessment design also demands cultivating AI literacy among students and educators, enabling them to critically evaluate AI outputs, identify limitations, and maintain epistemic agency in inquiry. Moreover, formative assessments supported by AI must protect student data privacy, comply with legal frameworks, and amplify learning rather than replacing instructional judgment. Professional development for teachers will be crucial to ensure they can interpret AI-supported insights, contextualize them in classroom practice, and intervene when AI feedback is misleading or incomplete. By grounding AI-supported assessments in evidence-centered design, rigorous fairness audits, and sustained human oversight, educators can ensure that technological innovations strengthen rather than compromise the validity, reliability, and inclusivity of science education

While acknowledging the potentials to science assessment, Chapter 10 also stresses the potential pitfalls of integrating AI in science assessments, particularly in three forms: cognitive outsourcing, bias, and issues of locality. Students may outsource cognitive effort by relying on AI to generate responses, undermining authentic engagement and reducing self-regulated learning opportunities, which over time can limit their ability to develop independent scientific reasoning and metacognitive skills. Teachers, too, risk over-reliance on AI for task design, scoring, and feedback, which could erode their assessment literacy, weaken pedagogical coherence, and diminish their professional agency in making context-sensitive judgments. Additionally, sociotechnical biases in AI scoring—stemming from historical inequities, skewed training datasets, or flawed algorithms—pose serious threats to fairness, disproportionately impacting marginalized learners such as English Language Learners, students from under-resourced schools, or those using non-standard dialects. Locality concerns further complicate adoption, as AI-supported assessments must align with national or regional standards, adapt to varied linguistic and cultural contexts, and remain accessible despite disparities in infrastructure, policy environments, and digital resources. Without careful attention, these challenges risk not only misrepresenting student ability but also reinforcing systemic inequities.

To ethically navigate these challenges, Chapter 10 suggests that educators and developers must reconceptualize assessment practices with human–AI collaboration at their core, ensuring that AI supplements rather than supplants teacher judgment and student engagement. This requires designing assessments that resist over-reliance on AI, such as multimodal and interactive tasks that preserve student agency, encourage critical thinking, and demand context-rich problem solving. Addressing bias involves collecting diverse, representative data across linguistic, cultural, and socioeconomic groups; adopting robust analytic rubric designs that foreground scientific reasoning rather than surface features; and applying statistical and sociological strategies for bias detection, auditing, and mitigation. Locality-sensitive design is equally vital: AI-supported assessments must respect cultural and linguistic diversity, integrate context-specific standards, and expand access through open-source models and equitable infrastructure so that assessment benefits are distributed fairly across global contexts. Transparency, frequent calibration of both human and machine scorers, and structured stakeholder involvement from teachers, students, and policymakers are essential to building trust and ensuring accountability. By embedding these practices—emphasizing fairness, explainability, data security, and ongoing human oversight—AI-supported science assessment can advance equity, authenticity, and educational value while respecting human dignity and promoting more just and inclusive science learning.

Taken together, these chapters highlight that AI is transforming science assessment into a more dynamic, responsive, and authentic practice while also surfacing new ethical and pedagogical challenges. AI enables performance-based, multimodal, and formative assessments that align more closely with the complexity of scientific reasoning and inquiry. At the same time, pitfalls such as cognitive outsourcing, sociotechnical bias, and misalignment with local contexts threaten the integrity and equity of assessment practices. Addressing these challenges requires robust frameworks grounded in validity, fairness, transparency, and human oversight. It also demands building AI literacy among educators and learners so that AI is approached not as a shortcut but as a partner in inquiry and evaluation. When guided by responsible and ethical principles, AI-supported science assessment can enhance both the rigor and inclusivity of science education, ensuring that innovations strengthen, rather than weaken, the educational enterprise.

### *5.5. Educational Outcomes*

*Chapter 11* highlights how AI has transformed the desired outcomes of science education by broadening its focus from factual knowledge and inquiry skills to include AI literacy, critical reflection, and ethical reasoning. Contemporary frameworks emphasize that students should not only understand scientific concepts but also know how AI functions within scientific inquiry, apply AI tools in authentic contexts, and critically evaluate AI-supported outputs. In this sense, science education now aims to cultivate competencies that align with the realities of an AI-driven world: collaboration with AI systems, recognition of algorithmic bias, and reflective judgment about the social, environmental, and epistemological implications of AI. These outcomes go beyond technical proficiency, preparing learners to navigate complex socio-scientific challenges and to exercise agency in shaping the role of AI in their personal, professional, and civic lives. Expanding on this transformation, the outcomes also incorporate dispositions such as adaptability, resilience in the face of technological uncertainty, and the ability to integrate interdisciplinary perspectives. For example, students are increasingly expected to connect their scientific reasoning with ethical, cultural, and political considerations, thereby situating AI not merely as a technical tool but as a socio-technical system with wide-ranging consequences. These expanded expectations indicate that science education is moving toward a holistic vision where cognitive, practical, and moral capacities are developed in parallel.

Addressing these new outcomes responsibly requires embedding ethical principles and reflective practices throughout science education. Ethical engagement entails teaching students to question the fairness, transparency, and accountability of AI systems, as well as to consider the implications of data bias, misinformation, and unequal access to digital resources. Reflection must be positioned as an ongoing practice, enabling students to move beyond surface-level judgments about accuracy and toward deeper inquiries into the assumptions, purposes, and societal impacts of AI technologies. To achieve this, educators need coherent learning progressions, scaffolded opportunities for inquiry, and curricula that balance technical skills with ethical reasoning. Responsible implementation also involves professional development for teachers so they can confidently integrate AI tools while guiding students through the complexities of ethical dilemmas. Schools and policy-makers must further ensure equitable access to AI resources, reducing digital divides that could exacerbate inequalities in learning outcomes. By fostering critical awareness and a sense of responsibility, science education can prepare students not merely to use AI, but to act as thoughtful stewards of its applications and consequences.

## 6. From Here Forward: With AI in Classrooms, What Roles Remain Uniquely Human?

As the preceding sections illustrate, AI is transforming nearly every dimension of science education—goals, procedures, materials, assessment, and outcomes. These transformations highlight AI's capacity to amplify efficiency, personalization, and authenticity in learning, while also raising questions of fairness, transparency, accountability, and integrity. Yet amid this sweeping change, a critical question remains: what roles in teaching and learning cannot, and should not, be delegated to machines? Identifying these uniquely human contributions is essential, not only to safeguard the integrity of education but also to ensure that AI functions as an enabler rather than a replacement of human judgment, creativity, and empathy. Without clarity on what remains distinctly human, there is a risk of drifting into educational models that privilege automation over meaning, standardization over equity, and efficiency over human growth. Thus, foregrounding the roles that remain uniquely human allows us to envision AI-supported classrooms that advance the transformative aims of science education rather than compromise them.

First, human educators remain indispensable as moral and relational anchors in classrooms. AI can generate solutions, feedback, or explanations, but it cannot genuinely care for students' well-being, cultivate belonging, or build trust across cultural and social divides. Teachers embody empathy, mentorship, and encouragement—qualities that sustain students' motivation, resilience, and identity development in ways no algorithm can replicate. For example, while AI-supported tutors can provide personalized feedback at scale, it is the teacher who notices when a student feels discouraged, recognizes subtle nonverbal cues, and responds with compassion. Beyond individual relationships, teachers also nurture classroom communities, cultivating shared norms of respect, inclusion, and collective inquiry. This relational dimension connects directly to earlier concerns about fairness and equity: without the human touch, AI risks reproducing existing disparities rather than addressing them. Teachers ground education in compassion and ethical responsibility, ensuring that science learning supports not just cognitive growth but also social and emotional flourishing.

Second, teachers and students alike bring interpretive and ethical judgment that AI lacks. Scientific reasoning often involves uncertainty, competing perspectives, and normative considerations. While AI systems may model probabilities or detect patterns, only humans can situate those results within cultural, ethical, and civic contexts. For example, AI can simulate climate change scenarios or analyze genetic data, but decisions about sustainability, equity, and justice require human deliberation. In this respect, the teacher's role extends beyond interpreting data to modeling ethical reasoning for students—demonstrating how to weigh evidence against moral principles, community needs, and long-term consequences. Likewise, students must be empowered to question AI outputs: Whose perspectives are reflected? What assumptions underlie the model? What values should guide our response? These interpretive and ethical functions connect directly to earlier discussions of accountability and transparency, ensuring that AI-supported science education remains socially responsible and aligned with democratic values.

Third, human agency is irreplaceable in fostering creativity, imagination, and curiosity. AI excels at recombining existing knowledge, yet it currently lacks the kinds of boundary-crossing insights and divergent thinking that spark true innovation. In classrooms, teachers inspire wonder, pose provocative questions, and create opportunities for students to pursue open-ended investigations. For instance, while AI might suggest multiple possible explanations for a phenomenon, it is the student's imaginative leap—the "what if" question—that drives new lines of inquiry. Teachers amplify this spirit by designing environments that value experimentation, embrace ambiguity, and legitimize failure as part of the discovery process. This aligns with earlier sections on goals and outcomes, where creativity, resilience, and interdisciplinary problem-solving were highlighted as essential competencies for an AI-driven future. Here, the uniquely human contribution lies not in efficiency but in nurturing the dispositions and curiosity that sustain lifelong learning and scientific innovation.

Ultimately, what remains uniquely human about AI-supported science education is the ability to co-construct meaning through dialogue and community. AI may simulate conversation, but it lacks the lived experience, vulnerability, and shared identity that underpin genuine dialogue. Classroom discussions, peer collaboration, and teacher-student exchanges cultivate epistemic humility and democratic participation, reminding learners that science is not only a body of knowledge but also a collective human endeavor. In dialogic interactions, misunderstandings become opportunities for learning, diverse perspectives enrich understanding, and students practice civic reasoning by negotiating differences respectfully. These human-centered practices connect to earlier warnings against over-reliance on AI and the risk of mechanizing science education. Dialogue and community remind us that knowledge is co-created, contested, and contextualized—an essential safeguard against the simplification or distortion of scientific reasoning by AI.

In this sense, the most vital human roles in AI-supported classrooms are not diminished by technology but amplified through intentional design. Teachers become orchestrators of inquiry, ethical stewards of technology use, and mentors of student identity and agency. Students, meanwhile, are not merely consumers of AI outputs but active participants who critique, extend, and reinterpret them. Thus, the integration of AI into science education calls not for a diminished view of humanity but for a sharpened appreciation of the qualities that remain distinctly human—empathy, judgment, creativity, and community. These qualities must remain at the heart of science education if AI is to truly serve the larger goals of equity, integrity, and human flourishing. Framed in this way, the partnership between humans and AI becomes less about efficiency and more about meaning, less about automation and more about agency, ensuring that the future of science education remains transformative, ethical, and unmistakably human.

**Declaration of AI Use**

Parts of this manuscript were prepared with the assistance of generative AI tools. The AI tools were used for language editing or paragraphing of the authors' writing. The authors reviewed, revised, and verified all AI-supported content for accuracy, appropriateness, and originality. No confidential or sensitive data were entered into AI systems. The final responsibility for the content rests solely with the authors.